\documentclass[prl,twocolumn]{revtex4}

\usepackage{graphicx}

\usepackage{dcolumn}

\usepackage{bm}

\usepackage{wasysym}

\usepackage{amssymb}

\def \be{\begin{equation*}}
\def \ee{\end{equation*}}

\begin{document}
\title{Helical Metal Inside a Topological Band Insulator}

\author{Ying Ran$^{1,2}$, Yi Zhang$^{1}$,  and Ashvin Vishwanath$^{1,2}$}

\affiliation{$^1$Department of Physics, University of California, Berkeley, CA 94720\\
$^2$Materials Sciences Division, Lawrence Berkeley National Laboratory, Berkeley, CA 94720}
\date{Printed \today}

\begin{abstract}
Topological defects, such as domain walls and vortices, have long
fascinated physicists. A novel twist is added in quantum systems
like the B-phase of superfluid helium He$_3$, where vortices are
associated with low energy excitations in the cores. Similarly,
cosmic strings may be tied to propagating fermion modes. Can
analogous phenomena occur in crystalline solids that host a plethora
of topological defects? Here we show that indeed dislocation lines
are associated with one dimensional fermionic excitations in a
`topological insulator', a novel band insulator believed to be
realized in the bulk material Bi$_{0.9}$Sb$_{0.1}$. In contrast to
fermionic excitations in a regular quantum wire, these modes are
topologically protected like the helical edge states of the quantum
spin-Hall insulator, and not scattered by disorder. Since
dislocations are ubiquitous in real materials, these excitations
could dominate spin and charge transport in topological insulators.
Our results provide a novel route to creating a potentially ideal
quantum wire in a bulk solid.
\end{abstract}

\maketitle

 Motivated by applications to spintronics, recent theoretical
work on the effect of spin orbit interactions on the band structure
of solids predicted the existence of novel `topological insulators'
(TIs) in two\cite{kane:226801,kane:146802,bernevig:106802} and three
dimensions\cite{fu:106803,moore:121306,roy-2006}. In two dimensional
TIs (or quantum spin-Hall insulators), in contrast to the gap in the
bulk, there are gapless modes at the edge which appear in time
reversed pairs with opposite spin and velocity. Transport
measurements on HgCdTe quantum
wells\cite{bernevig:Science,molenkamp:spin-hall} have provided
evidence for the existence of these `helical' edge
states\cite{kane:226801,helical1,helical2}. In three dimensions,
topological insulators are classified as strong ($\nu_0=1$) or weak
($\nu_0=0$), the former has surface states with an odd number of
Dirac points. In addition, 3D TIs are characterized by indices
$(\nu_1,\,\nu_2,\,\nu_3)$ with respect to a basis of reciprocal
lattice vectors $(\vec{G}_1,\,\vec{G}_2,\,\vec{G}_3)$. They define a
time reversal invariant momentum (TRIM) $\vec{M}_\nu = \frac12
(\nu_1 \vec{G}_1+\nu_2 \vec{G}_2+\nu_3 \vec{G}_3)$. The recent
prediction \cite{fu:106803} and evidence from angle resolved
photoemission experiments \cite{hasan:Nature} for a strong TI phase
in the alloy Bi$_{0.9}$Sb$_{0.1}$, has led to heightened interest in
these systems.

\begin{figure}
\includegraphics[width=0.35\textwidth]{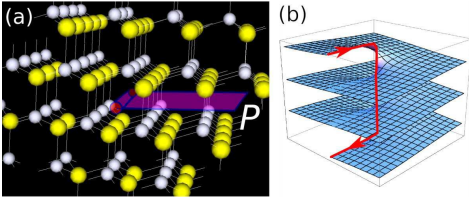}
\caption{(color online) [a] Diamond lattice screw dislocation
along $\vec{B}=a(1,1,0)$ direction (shown as a cylinder with red
ends). The two diamond sublattices are shown as yellow and white
spheres. A cutting plane P that was used to generate the
dislocation is shown, orthogonal to the strength $t$ bond along
$a/2(1,-1,-1)$. [b] Screw dislocation in a stacked 2D topological
insulator. Unpaired edge modes (one member of the edge mode pair is
shown in red) of the top and bottom layer must propagate through the
dislocation by continuity.} \label{fig1}
\end{figure}

{\em Dislocations} are line defects of the three dimensional
crystalline order, characterized by  a vector $\vec{B}$ (Burgers
vector), which is a {\em lattice vector}. It must remain constant
over the entire dislocation. The center of the dislocation is at
$\vec{R}(\sigma)$ where $\sigma$ parameterizes the line defect.  A
convenient way to visualize a dislocation is via the Volterra
process. One begins with the perfect crystal and chooses a plane $P$
that terminates along the curve $\vec{R}(\sigma)$ where the defect
is to be produced. The crystal on one side of the plane  $P$ is then
displaced by the lattice vector $\vec{B}$, and additional atoms are
inserted or removed if required. At the end, crystalline order is
restored everywhere except near the curve $\vec{R}(\sigma)$. A screw
dislocation (see Figure \ref{fig1}) is a line defect with its tangent vector
$\vec t=d\vec{R(\sigma)}/d\sigma\parallel \vec{B}$
, while an edge dislocation is also a line defect with $\vec{t}\perp \vec{B}$. In general, a
dislocation varies between these two simple types along its length.

We now state the main result of this paper. Consider a dislocation
with Burgers vector $\vec{B}$ inside a three dimensional topological
insulator characterized by $(\nu_0;\, \vec{M}_\nu)$. Iff
\begin{equation}
\vec{B}\cdot \vec{M}_{\nu} = \pi \, ({\rm mod} \, 2\pi)
\label{Eq:main}
\end{equation}
the dislocation induces a pair of one dimensional modes bound to it
that traverse the bulk band gap. The modes are related by the
Kramers time reversal symmetry transformation and reflect the
nontrivial topology of the band structure. We shall refer to this
one dimensional state as the `helical metal'. While a trivial
insulator could also develop one dimensional propagating modes along
a dislocation, the precise count obtained here - a single Kramers
pair - cannot arise. The topological stability of the helical metal
results from time reversal symmetry, which prohibits backscattering
between the oppositely propagating modes\cite{kane:146802}. From
Eqn. \ref{Eq:main} we note that some dislocations e.g. with $\vec{B}
\perp \vec{M}_\nu$ do not carry these helical modes. Also, there is
the 'pristine' strong topological insulator $\nu_0=1,\,
\vec{M}_\nu=0$, where none of the dislocations induce helical modes.
Experiments indicate that Bi$_{0.9}$Sb$_{0.1}$ is a strong TI with a
nontrivial $\vec{M}_\nu$.

{\bf Dislocations in the Diamond Lattice TI }  Instead of plunging
into a general proof of our results, we first present a numerical
calculation of a dislocation in a simple TI model on the diamond
lattice\cite{fu:106803}:
\begin{eqnarray}
H = t \sum_{\langle ij\rangle} c^\dagger_{i\sigma} c_{j\sigma}
+i\frac{8\lambda_{SO}}{(2a)^2}\sum_{\langle\langle
ik\rangle\rangle}c^\dagger_{i\sigma}(\vec{d}^{\,1}_{ik}\times
\vec{d}^{\,2}_{ik})\cdot \vec{\sigma}_{\sigma\sigma'} c_{k\sigma'}
\label{Eq:FuKaneDiamond}
\end{eqnarray}
where $\vec{d}^{\,1}_{ik},\, \vec{d}^{\,2}_{ik}$ are the two nearest
neighbor bond vectors leading from site $i$ to $k$, 
$\vec{\sigma}$ are the spin Pauli matrices, and $2a$ is the cubic cell size. The system is a Dirac
metal at this point, and a full gap is opened upon distorting the
lattice, by making the nearest neighbor hopping strengths $t+\delta
t$ along one of the directions e.g. $\frac{a}2 (1,1,1)$. If $\delta
t>0$ ($\delta t<0$), the system is a strong (weak) topological
insulator with $\nu=1$ ($\nu=0$) and
$\vec{M}_\nu=\frac{\pi}{2a}(1,1,1)$. We now study a screw
dislocation and use periodic boundary conditions - which requires us
to consider a separated pair of dislocations. Translation symmetry
is present along the dislocation axis, and we label states by $k$,
the crystal momentum in this direction. The precise geometry we use
is shown in Figure \ref{fig1}a. If $\vec{a}_1=a(0,1,1)$,
$\vec{a}_2=a(1,0,1)$, $\vec{a}_3=a(1,1,0)$ are the basis vectors,
the dislocations are along the $\vec{a}_3$ axis.

\begin{figure}
\includegraphics[width=0.45\textwidth]{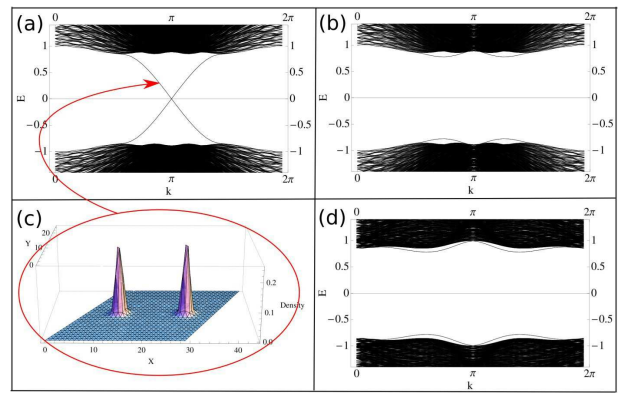}
\caption{(color online) Electronic spectrum of the diamond lattice strong TI
($\vec{M}_\nu=\frac{\pi}{2a}(1,1,1)$) in the presence of a pair of
screw dislocations. Electronic states are shown as a function
of $k$, the wavevector along the dislocation. For [a][b][c], the dislocations are oriented along the
$\vec{a}_3=a(1,1,0)$ axis. [a] Spectrum, when the
dislocation Burgers vector $\vec{B}=\vec{a}_3$ and satisfies
$\vec{B}\cdot\vec{M}_\nu=\pi$. A pair of counter-propagating modes
per dislocation is found, that span the band gap (each midgap level
happens to be doubly degenerate). [b] Spectrum when
$\vec{B}=2\vec{a}_3$ and $\vec{B}\cdot\vec{M}_\nu=0 {\rm
mod}\,2\pi$. No dislocation modes occur. [c] Probability
distribution of a pair of midgap modes shown in the
$\vec{a}_1,\,\vec{a}_2$ plane that intersects the dislocations. The
modes are bound near the dislocations. [d] The dislocations are oriented along $\vec{B}=a(0,1,-1)$ and $\vec{B}\cdot\vec{M}_\nu=0$. No dislocation modes occur. Calculations were done on a 36x36x18 unit cell
system with periodic boundary condition along
$\vec{a}_1,\,\vec{a}_2,\,\vec{B}$ directions, and dislocations
separated by a half system size. The hopping parameters used in
Eqn.\ref{Eq:FuKaneDiamond} are $t=1$, $t+\delta t=2$, and
$\lambda_{SO}=0.125$.}
\label{fig2}
\end{figure}

For a pair of unit screw dislocations with $\vec{B}=a(1,1,0)$, we find four one dimensional modes, two down moving and
two up moving modes, within the bulk gap (Figure \ref{fig2}a). The
wavefunctions of these modes are peaked along the two dislocations
(Figure \ref{fig2}c) - hence a pair of oppositely propagating modes
is present for each dislocation line. The modes are time reversal
(Kramers) conjugates of each other.
Not all dislocations carry these gapless
modes - if we double the Burgers vector of this dislocation (Figure
\ref{fig2}b), or consider a different relative orientation when the
unequal bond is orthogonal to the Burgers vector (Figure \ref{fig2}d),
the one dimensional modes do not appear. The sign of $\delta t$ is immaterial - both weak and strong topological insulators
show this physics.

Note the condition for a gapless line mode to exist on
a dislocation {\em cannot} depend on its orientation, since the modes must
propagate through the entire defect. Hence, it can only depend on
the Burgers vector $\vec{B}$. Therefore, although we have
explicitly considered the case of a screw dislocation, the result
also holds for an edge or mixed dislocation with the same Burgers
vector.

We now present an analytical rationalization for the existence of
gapless modes along the dislocation line, in this model of the
strong TI. We study the model in a limit that is analytically
tractable, but nevertheless adiabatically connected to the parameter
regime of interest. Consider the imaginary plane $P$ (orthogonal to
$(1,-1,-1)$), used to construct the dislocation, which ends along
the dislocation line. Extend this plane through the entire crystal,
weakening all bonds that are cut by it. Now, the crystal is almost
disconnected into two disjoint pieces. Gapless states are generated
on the two surfaces $S_+,\,S_-$, each with a single Dirac node
located at the surface momentum $\vec{m}_D=\frac{\pi}{2a}(1,1,0)$.
Electronic states in the vicinity of the Dirac point can be expanded
as $\Psi(x_1,x_2)\sim e^{i\vec{m}_D\cdot\vec{x}}\psi(\vec{x})$,
where $\psi(\vec{x})$ varies slowly over a lattice spacing.  The
effective Hamiltonian for such states on the two surfaces is $\pm
H_0=p_1\sigma_x+p_2\sigma_y$ \cite{fu:106803,fu:096407}, where
$p_1$, $p_2$ are surface momenta measured from the Dirac point, with
$p_1$ along the dislocation line (the $(1,1,0)$ direction). Thus if
$\mu_z=\pm 1$ represents the two surfaces, the effective Hamiltonian
is $H=H_+(p)\mu_z$. Now, if we imagine reconnecting the surfaces
{\em without} creating a dislocation, we add a hopping term between
the two surfaces: $H=H_+(p)\mu_z+m\mu_x$, which leads to fully
gapped dispersion, with gap $2m$, which can be adiabatically
connected to the uniform bulk insulator.

Now consider introducing a screw dislocation, by displacing the
surface to one side of the plane $P$  by the lattice vector
$\vec{B}=a(1,1,0)$. In our surface coordinate system, the
dislocation line is along the $x_1$ axis, at $x_2=0$, and the plane
$P$ is at $x_2>0$. The crucial observation is that when reconnecting
the bonds across the plane $P$, if the Dirac node is at momentum
$\vec m_D$ with $\vec{m}_D\cdot\vec{B}=\pi$, then the Dirac
particles acquire a {\em phase shift} of $\pi$. The effective
Hamiltonian in the presence of a dislocation is then: $ H_{\rm
dis}=H_+(-i\hbar\nabla)\mu_z+m(x_2)\mu_x $ where the Dirac mass term
is: $m(x_2<0)=m$ but $m(x_2>0)=-m$. Such Dirac Hamiltonians are well
known to lead to low energy modes\cite{CallanHarvey} that are
protected by index theorems. At $p_1=0$, a pair of midgap modes are
present $ \psi_{\pm}(\vec x)=e^{\frac{1}{\hbar v_2}\int_0^{x_2} dy
m(y)}\psi_{0,\pm} $ where $\psi_{0,\pm}$ are the two solutions of:
$\mu_y \sigma_y \psi_0 =\psi_0$. For finite $p_1$ these split to
give rise to the left and right moving modes.

More generally, if there are several such surface Dirac modes
$\vec{m}^{i}_D$, what is required is that an {\em odd} number of
them acquire a $\pi$ phase shift on circling the dislocation $\sum_i
\vec{m}^{i}_D\cdot \vec{B}=\pi ({\rm mod}\,2\pi)$. This guarantees
the existence of at least a pair of gapless 1D modes on the
dislocation. If on the other hand, $\sum_i \vec{m}^{i}_D\cdot
\vec{B}=0 ({\rm mod}\,2\pi)$, there are no protected modes on the
dislocation.

{\bf Dislocations in a General TI} Armed with these insights for a
specific model, one may discuss the general case of an arbitrary
dislocation in a topological band insulator. The condition for the
existence of a protected 1 D mode is given by equation
\ref{Eq:main}, and derived in the general case in the paragraph
below. For the weak TI case, however, a more intuitive derivation of
this result is possible. The weak TI is adiabatically connected to a
stack of decoupled two dimensional TIs, stacked along the
$\vec{M}_\nu$ direction. The top and bottom surfaces of the stack
are fully gapped. Consider creating a screw dislocation in such a
stack, by cutting and regluing layers. Cutting a plane results in a
pair of edge modes, which are gapped after the planes are glued back
together. However, edge modes in the top and bottom layers are left
out in this process (see Figure \ref{fig1}b). Since a single pair of such modes cannot begin
or end, the helical mode must propagate through the dislocation core
to complete the circuit. A similar result holds for three
dimensional Chern insulators \cite{PhysRevB.45.13488}, characterized
by  a reciprocal lattice vector $\vec{G}_0$. There, a dislocation
line should have $\frac{\vec{G}_0\cdot\vec{B}}{2\pi}$ chiral modes
propagating along it.

We now derive the main result equation \ref{Eq:main}. First, let us
briefly review the meaning of the topological insulator invariants
$(\nu_0;\,\vec{M}_\nu)$. These are conveniently expressed in terms
of quantities $\delta_{i=(n_1 n_2 n_3)}$ with $n_i=0,1$ defined at
the 8 TRIMs of the 3D Brillouin zone(BZ) $\Gamma_{i=(n_1 n_2 n_3)} =
(n_1 \vec G_1 + n_2 \vec G_2 + n_3 \vec G_3)/2$, which can be
computed for a given band structure \cite{fu:106803}. The gauge
invariant indices are given by: $(-1)^{\nu_0} = \prod_{n_j = 0,1}
\delta_{n_1n_2n_3}$ and $ (-1)^{\nu_{i=1,2,3}} = \prod_{n_{j\ne i} =
0,1; n_i = 1} \delta_{n_1 n_2 n_3}$.\cite{fu:106803} The indices
$\nu_i$ carry information about the surface band structure, in
particular, the number of times that the surface bands cross a
generic Fermi energy which lies within the bulk gap, along a line in
the surface BZ. For example consider the surface spanned by the
basis vectors  $\vec a_1,\vec a_2$ normal to $\vec G_3$. The parity
(even vs odd) of the number of such band crossings  $N_{\rm cross}$,
when connecting two surface TRIMs $\vec{m}_1,\,\vec{m}_2$ is a
topologically protected quantity \cite{fu:106803}:
\begin{equation}
(-1)^{N_{\rm
cross}}=\delta_{\vec{m}_1}\delta_{\vec{m}_2}\delta_{\vec{m}_1+\frac{\vec{G}_3}2}\delta_{\vec{m}_2+\frac{\vec{G}_3}2}.
\label{Ncross}
\end{equation}
Thus $\nu_1=N_{\rm cross} ({\rm mod } \, 2)$ between
$\vec{m}_1=\frac{\vec G_1}{2}$ and $\vec{m}_2=\frac{\vec G_1+\vec
G_2}{2}$ on this surface.

Without loss of generality, consider now creating a screw
dislocation in a TI with Burgers vector $\vec{B}=\vec a_1$. This is
created using the Volterra procedure by a cutting $\vec a_1-\vec
a_2$ surface containing the dislocation. Atoms on one side of the
surface are displaced by $\vec{B}$ and reconnected. The crucial step
in our analytical derivation of the helical modes was whether there
are even or odd number of surface Dirac nodes that acquire a $\pi$
phase shift on creating the dislocation. Although in general, the
surface band structure may be quite complicated, for the purpose of
deriving robust topological properties one may adiabatically deform
the crystal structure (during which the bulk gap remains open, {\em
and} the Bravais lattice structure stays fixed) so that the surface
modes are always Dirac nodes that cross the band gap, centered at
the surface TRIMs. Then, there are two surface TRIMs
$\vec{m}=\{\frac{\vec{G}_1}{2},\,\frac{\vec{G}_1+\vec{G}_2}{2}\}$
for which $\vec{B}\cdot\vec{m}=\pi ({\rm mod} \, 2\pi)$, i.e.
acquire a $\pi$ phase shift on creating the dislocation. If the
number of surface Dirac nodes present in total at these two TRIMs
$N_{\rm Dirac}$ is odd, then the dislocation will host a protected
helical mode. Clearly, the parity of $N_{\rm Dirac}$ is the same as
the number of band crossings between the two surface TRIMs given by
Equation \ref{Ncross}. Thus, the condition for the helical modes is
$\nu_1=1$ or $\vec{M}_{\nu}\cdot\vec{B}=\pi({\rm mod} \, 2\pi)$.

In general, the Burgers vector is a multiple of the primitive
lattice vector. If it an odd multiple, the derivation above is
unaffected. If it is an even multiple, there are no dislocation
modes since the phase shifts are always trivial. Hence we arrive at
\ref{Eq:main}, for the existence of protected helical modes along
the dislocation.

{\bf Effect of Disorder} It is sometimes stated that in the presence
of disorder, only the $\nu_0$ index is robust, while the
$\vec{M}_\nu$ index is irrelevant. Hence weak TIs are believed to be
unstable in the presence of disorder \cite{fu:106803}. Since our
results are controlled precisely by the $\vec{M}_\nu$ index, they
shed light on when they retain meaning. If the disorder is so strong
that the dislocations are no longer well defined, then the only
remaining distinction is the $\nu_0$ index. Similarly, an insulator
with $\vec{M}_\nu\neq 0$ can be converted to one with
$\vec{M}_\nu=0$, by introducing a potential modulation with period
$\vec{M}_\nu$. Elementary dislocations that carry helical modes will
then cost infinite energy per unit length. However, for a
sufficiently weak disorder potential, the dislocations are expected
to remain well defined, in which case the distinction captured by
$\vec{M}_\nu$ is physically relevant. Certainly, in real crystals
which are inevitably disordered, dislocations are well defined
objects, and hence the physics discussed in this paper should apply.

\begin{figure}
\includegraphics[width=0.49\textwidth]{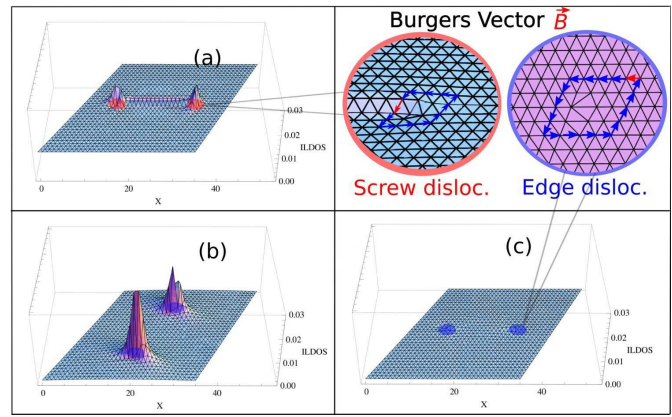}
\caption{(color online) Surface LDOS as measured by STM, for various
dislocation configurations, in the diamond lattice strong TI with
parameters as in Fig.2. In all cases dislocation pairs are separated
by a half system size, and are directed along $\vec{a}_3$.
Topography of the integrated LDOS in the energy window $[-0.1,0.1]$
is shown, with insets displaying energy dependent LDOS at special
points on the surface. [a] The $(111)'$ surface intersecting a pair
of screw dislocations with $\vec{B}=a(1,1,0)$. These carry 1D modes
which give rise to a peak in the integrated LDOS. In [b] and [c],
edge dislocations on the $(\bar{1}\bar{1}1)$ surface. In [b], the
dislocation satisfies $\vec{B}\cdot\vec{M}_\nu=\pi$, and has 1D
modes visible in the LDOS while in [c] the dislocation has
$\vec{B}\cdot\vec{M}_\nu=0$, hence no 1D line mode and no enhanced
LDOS.In [a][b][c] only the top layer of atoms is shown, and
calculations are done on 36x36x18 unit cell system, with periodic
boundary conditions in the $\vec{a}_1,\,\vec{a}_2$ directions.}
\label{fig4}
\end{figure}

{\bf Experimental Consequences} In real solids, dislocations are always present and the predicted
helical modes should have important experimental consequences.
Scanning tunneling microscopy (STM) of a topological insulator
surface where dislocations terminate, is particularly well suited to
verify these predictions. Since the precise atomic arrangement is
visualized by STM, the nature of the dislocations involved can be
characterized. At the same time, the finite density of states
associated with the one dimensional modes will lead to an enhanced
tunneling density of states near the dislocation core. This will be
particularly striking if one tunes to an energy where the surface
modes contribute minimally (eg. at the Dirac node). Such experiments
are immediately feasible on the putative topological insulator
Bi$_{0.9}$Sn$_{0.1}$ with A7 structure. Since the band structure of
that material is complex and still debated \cite{teo:045426}, we
demonstrate qualitatively what is expected, in the simpler diamond
lattice strong TI model \ref{Eq:FuKaneDiamond}. Moreover, since the
crystal symmetry in the two cases share several features - such as a
three fold axis parallel to the $\vec{M}_\nu$ vector, some
predictions for the diamond lattice model may be directly relevant
to the A7 structure. First consider the surface orthogonal to the
strong bond $(1,1,1)$ direction. We choose this surface to cut three
strength $t$ bonds (rather than a single strong bond)in order to
obtain a surface band structure similar in some respects to the
$(111)$ surface of Bi$_{0.9}$Sn$_{0.1}$. We call this the $(111)'$
surface. This surface has a single Dirac node centered at the
$\Gamma$ point.
We consider a pair of separated screw dislocations with
$\vec{B}=a(1,1,0)$, which carry 1D helical modes, that terminate on
the surface. STM measures the local density of states (LDOS) at a
given energy and surface location $\rho(\vec{r},E) = \sum_\alpha
|\psi_\alpha(\vec{r})|^2 \delta(E-E_\alpha)$, where the $\alpha$ sum
runs over all eigenstates. The dislocations appear as peaks in the
LDOS as shown in Figure \ref{fig4}a, due to these helical
modes. Note, the the ledge connecting the two screw dislocations on
the surface also shows an enhanced DOS.

More striking evidence for the relation \ref{Eq:main} appears when
we consider the edge dislocations on the $(\bar{1}\bar{1}1)$
surface. If the Burgers vector of the dislocation is
$\vec{B}=a(1,0,1)$ the one dimensional modes are present and visible
in the LDOS (Figure \ref{fig4}b), but if $\vec{B}=a(1,-1,0)$ then
they are absent (Figure \ref{fig4}c). Here, there is no physical
line connecting the two defects.

Since elastic scattering by nonmagnetic impurities cannot scatter
electrons between the counter-propagating helical modes of the
dislocation, they behave as ideal quantum wires which may be relevant for spin and charge transport applications
at low temperatures when inelastic processes are frozen out. The challenge will
be to separate this conduction mechanism from protected surface mode
conduction, which is also present. In conduction across the short
direction of anisotropic samples (eg. disc shaped), dislocation
induced direct conduction paths should dominate over surface
conduction. If dislocations can be induced in a controlled fashion
during the growth process, then the direction dependent condition
for the existence of helical modes, and dislocation density
dependence of conductivity can be used to isolate this contribution.
Another possibility is to introduce magnetic impurities on the
surface of the insulator, which could potentially localize surface
modes but have little impact on dislocation helical modes deep in
the bulk. A rough estimate of the excess electric conductivity induced by
dislocation modes for a dislocation density $n_d\sim 10^{12}{\rm
m}^{-2}$ and a low temperature scattering length $l\sim 1\mu m$
along the dislocation\cite{molenkamp:spin-hall} yields $\rho =
\frac{h}{2e^2}\frac1{n_dl}\sim 10{m}\Omega {\rm m}$, which should be
experimentally accessible.

Funding from NSF DMR-0645691 is acknowledged.


\bibliographystyle{apsrev}

\bibliography{/home/ranying/downloads/reference/simplifiedying}

\end{document}